# Exploiting SNMP-MIB Data to Detect Network Anomalies using Machine Learning Techniques

Ghazi Al-Naymat
Computer Science Department
Princess Sumaya University for Technology
Amman, Jordan
g.naymat@psut.edu.jo

Mouhammd Al-kasassbeh, Eshraq Al-Hawari
Computer Science Department
Mutah University
Karak, Jordan
mouhammd.alkasassbeh@mutah.edu.jo, eshraq.alhawari@gmail.com

*Abstract*—The exponential increase in the number of malicious threats on computer networks and Internet services due to a large number of attacks makes the network security at continuous risk. One of the most prevalent network attacks that threaten networks is Denial of Service (DoS) flooding attack. DoS attacks have recently become the most attractive type of attacks to attackers and these have posed devastating threats to network services. So, there is a need for effective approaches, which can efficiently detect any intrusion in the network. This paper presents an efficient mechanism for network attacks detection and types of attack classification using the Management Information Base (MIB) database associated with the Simple Network Management Protocol (SNMP) through machine learning techniques. This paper also investigates the impact of SNMP-MIB data on network anomalies detection. Three classifiers, namely, Random Forest, AdaboostM1 and MLP are used to build the detection model. The use of different classifiers presents a comprehensive study on the effectiveness of SNMP-MIB data in detecting different types of attack. Empirical results show that the machine learning techniques were quite successful in detecting and classifying the attacks with a high detection rate.

## I. Introduction

The rapid development of the Internet and the growing use of wired and wireless networks have increased the security breaches and malicious attacks. Recently, a variety of network attacks have posed devastating threats to network resources and its security. Some of these attacks are launched to make network services unavailable, such as flooding attacks (e.g., DoS/Distributed DoS and Internet Worm attacks). Other attacks are used to obtain unauthorized access (e.g., buffer overflow and brute force attacks) [1]. Among all types of attack, the DoS/DDoS attacks are considered the most significant and dangerous attacks [26]. Moore et al. [2] reported that these types of attacks are the main threat to the Internet, and 90-94% of them are deployed using TCP [1]. Therefore, it is essential to have rapid detection techniques to detect these attacks with high detection rate for secure and more reliable network services [3]. Intrusion detection (ID) is a key component of any security mechanism. ID aims to protect network security by detecting any abnormal activity occurring in a computer system or network by comparing it with a normal event to identify signs of intrusion [4]. ID is being implemented in two general approaches: Anomaly Intrusion Detection (AID) and Misuse Intrusion Detection (MID) [5].

AID constructs a normal profile of network behavior to identify any deviations in normal behavior profile as a possible result of attack activities. AID is based on assumption that the intruder behavior will be different from the legitimate user behavior, so it is useful to detect new types of attack. However, MID used pattern matching to detect intrusions depending on signatures of known attacks to detect it with high accuracy [6]. Intrusion detection systems (IDSs) use various data records, collected from target computer or network, to examine them for detecting network attacks [4]. A key aspect of any IDS is the type of the used dataset, so obtaining the right type of data about network traffic is essential for accurate and fast intrusion detection [7]. Most of the current research in the field of network intrusion detection depends on analyzing raw packet data to evaluate the security status of computer systems and networks, this will lead to a significant processing burden and late detection time [1], [6]. Hence, other data sources about network traffic have been provided by several network management protocols, such as CIMP (Common Management Information Protocol), RMON (Remote Network Monitoring), and SNMP (Simple Network Management Protocol). It should be known that the most deployed and widely used protocol is SNMP [8].

SNMP is a popular protocol for network management. It collects information from different types of network devices, such as routers, switches, and hubs on an Internet Protocol (IP) network. SNMP offers powerful statistical information about what is happening in network devices through a database of management variables called Management Information Base (MIB) [1], [8], [9]. Fig. 1 illustrates that SNMP and MIB data are fundamental elements of the general Internet management model. The MIB variables provide a numerous traffic infor-mation at different layers and protocols: IP, ICMP, TCP, UDP, etc. The collected information from the network devices can be passively monitored and could be used to characterize network behavior and, therefore, can be used for network intrusion detection [5]. By utilizing the fine grained data provided from SNMP-MIB in intrusion detection, some of the challenges in network intrusion detection can be avoided, and IDS promises a lower processing overhead analysis with high flexibility of deployment [1], [9], [10]. This is what motivates the work presented in this paper, which attempts to present a mechanism for anomaly based intrusion detection using SNMP-MIB dataset. Various techniques have been proposed and applied for network anomaly detection using SNMP-MIB data, such as statistical analysis and machine learning methods, which have been extensively used as anomaly detection techniques [11]. Examples of these approaches include SVM [1], [12], decision trees [6], [13], ANN [14] and K-mean clustering [11], to mention a few. The motivation behind this research is set as follows: (1) Due to the importance of IDS in any network environment and the availability of rich SNMP-MIB dataset. (2) Given that some attacks, such as Slowloris and Slowpost, have not been used in previous research. (3) To examine the validity and effectiveness of the use of SNMP-MIB data to classify up to date network attacks.

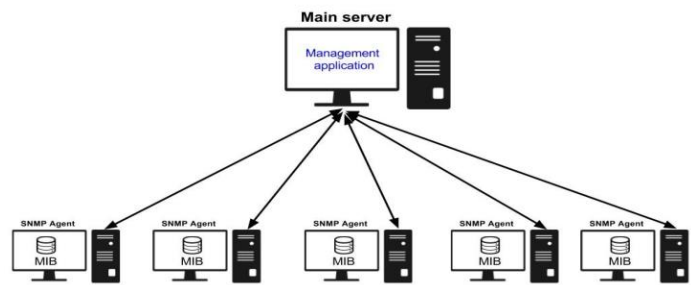

Fig. 1. SNMP-MIB and the general Internet management model.

In this paper, we make the following contributions:

1) We present a MIB based mechanism for network attacks detection and attacks classification using machine learning techniques.
2) We use a recent SNMP-MIB dataset generated by [15] to effectively detect DoS and brute force attacks.
3) We used three different classifiers (AdaboostM1, Random forest and MLP classifiers) for detecting and classifying the DoS attacks (TCP-SYN flooding, UDP flooding, ICMP-ECHO flooding, HTTP flood, Slowloris, Slowpost) and Brute Force attack.
4) The proposed model shows high accuracy and good detection rates.
5) The results demonstrate that using SNMP-MIB data with machine learning techniques is a very effective approach for the detection of DoS and brute force attacks.

The rest of the paper is organized as follows: Section II includes an overview of related works. Section III presents the general architecture of our method for attacks detection and classification based on SNMP-MIB data. Section IV describes the experiment implementation. In Section V, we discuss the results. Finally in Section VI, we present our conclusions.

## II. Related Work

Many researchers have extensively studied anomalies and attacks detection in computer networks during the last decade. Surveys about anomalies detection were presented and many different approaches have been proposed and implemented [6]. The vast majority of the solutions presented so far in the scope of network anomalies and attacks detection have been focused on analyzing raw traffic features such as a number of packets, IP addresses, ports, network flow, etc.). On the other hand, other solutions based on SNMP-MIB data as data source have been proposed for network anomaly detection.

Many studies have exploited SNMP-MIB data in the field of network anomalies detection. Some Researchers presented approaches based on statistical analysis of the MIB data, while others have recently utilized machine learning techniques to detect network attacks and other anomalies. The first attempt to exploit SNMP for network security is performed by Cabrera et al. [16]. They proposed a methodology for the early detection of Distributed Denial of Service (DDoS) attacks by applying statistical tests for causality to extract MIB variables that contain precursors to attacks. They used 91 MIB traffic variables from five groups (IP, ICMP, TCP, UDP and SNMP). Three types of DDoS attack (Ping Flood, Targa3 and UDP Flood) were conducted on a research test-bed with controlled loads in traffic. Their work has shown that it is possible to extract a precursor to a DDoS attack using MIB traffic variables and to detect these attacks before the target is shut down with about 1% rate of false alarms.

Yu et al. [1] and Bao. [6] utilized a machine learning approach based on a Support Vector Machine (SVM) for network intrusion detection based on SNMP-MIB data. They gathered 13 SNMP-MIB variables corresponding to 4 MIB groups (IP, ICMP, TCP, and UDP). The proposed system was constructed in a hierarchical SVM- based structure for attack traffic detection and classification into different types of attack (TCP-SYN flood, UDP flood and ICMP flood). They concluded that they had achieved fast detection with high detection accuracy 99.27% and with a low rate of false alarms using SVM and the key MIB variables that had been selected from a correlation feature selection mechanism (CFS).

An extended architecture of the system in [1], [6] was proposed by Yu et al. [9]. A system based on C4.5 algorithm with two level hierarchical structure was performed to detect traffic flooding attacks and classify them into different types of attack (TCP-SYN flood, UDP flood and ICMP flood). In offline mode, they performed classification and association rule mining facilitated by the C4.5 algorithm, while in the online mode they collected SNMP-MIB data and detected DoS/DDoS attacks, subsequently passing the result to the detection module. They reported that attack detection accuracy was about 99.13%.

Hsiao et al. [17] constructed a detection model based on applying Decision Tree (C4.5), Nave Bayesian and Support Vector Machine (SVM) data mining techniques that use SNMP-MIB data for Address Resolution Protocol (ARP) spoofing attack detection. They evaluated the performance of the proposed model using 6 MIB variables from the Interface group, and their results demonstrated that Decision Tree (C4.5) and SVM produce a better performance by accuracy rate, false alarm rate and missing rate than Nave Bayesian, while the C4.5 algorithm achieved the highest accuracy rate of about 95.9%.

In [13], Namvarasl and Ahmad Zadeh presented an intrusion detection system based on SNMP-MIB and machine learning. Their system consisted of three modules: the first for selecting key MIB variables from three classification algorithms (C4.5, RIPPER and attribute selection), and the second for generating an intrusion detection model based on the chosen variables and detecting DoS/DDoS attacks in real time in the third module. The dataset used in their system consisted of appropriate MIB variables among 66 variables corresponding to 4 MIB groups (IP, ICMP, TCP, and UDP) and involving a TCP-SYN flood attack, UDP flood attack and ICMP flood attack. Finally, they tested their proposed system and achieved about 99.03% accuracy rate by using a neural network algorithm among three classification algorithms (Neural network, Bayesian network and C4.5).

Cerroni et al. [18], [19] presented a decentralized system consisting of a peer-to-peer network of monitoring stations to collect SNMP-MIB statistical data and analyze them using distributed data mining techniques. In [18], authors introduced new supervised distributed classification algorithms (Distributed AdaBoosM1-MultiModel and Distributed AdaBoostM1-SingleModel) for the purpose of network attack detection and classification. The new classification algorithms have been evaluated and tested for decentralized environments using statistical SNMP-MIB data that were gathered locally from each device throughout the experiment. They collected 14 SNMP variables related to IP and TCP protocols involving different types of attack, including DoS, DDoS, TCP Port Scanning, SSH Denial of Service and SSH Brute Force attack. Finally, they concluded that the experimental results of the distributed classification algorithms achieved accuracy in the classification of network attacks.

Cerroni et al. [19] used unsupervised distributed data clustering techniques based on a K-means algorithm provided by the WEKA tool. They used 14 SNMP variables from the TCP group selected by using the correlation-based feature selection algorithm among SNMP data collected at regular intervals in the experiment. The efficiency of the algorithm was tested for several types of attack, including Denial of Service, Distributed Denial of Service, Denial of Service on SSH and Brute Force on SSH attack.

Another approach relied on machine learning and data mining techniques for attack detection and classification using SNMP-MIB data was proposed by Priya et al. [20]. They proposed a system named Protocol Independent Detection and Classification (PIDC) to detect and classify Distributed Reflection Denial of Service (DRDoS) attacks, such as a DNS attack and a TCP SYN reflection flooding attack. They captured 13 MIB variables from the TCP and DNS groups, and they used a rank correlation-based detection algorithm to determine the relationship between these variables. The C4.5 classification algorithm was then used with MIB variables to classify the type of attacks that were considered in this research. Their method achieved about 99% true positive rate and 1% false positive rate in the detection of reflected attacks.

Al-Kasassbeh [21] has adopted the distributed model in order to exclude the scalability problems in the network. His work showed that the statistical methods based on the Wiener filter that upgraded to the mobile agent could be used to detect the abnormality attempts. He took advantage of the correlation matrix between the input MIB variables and the cross-correlation with the desired MIB variables to detect abnormal situations. Al-Kasassbeh used only limited number of MIB variables.

In [22], Al-Kasassbeh applied statistical methods based on the Wiener filter combined with mobile agent technology to detect anomalies in the network traffic by using a set of MIB variables from 2 MIBgroups (interface and IP). The presented algorithm was tested against four kinds of network attacks (Buffer overflow attack, Decoy port-scan attack, Brute force attack and Null session attack) to show the effectiveness of the algorithm in network intrusion detection.

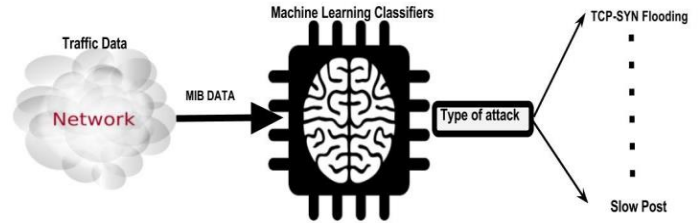

Fig. 2.  The overall architecture of the detection framework.

Some other studies utilized data obtained from SNMP-MIB for the purpose of intrusion detection in wireless networks [23], [24]. In [23], authors proposed a multi agent based intrusion detection system to detect the intrusion locally in mobile wireless networks using information from SNMP-MIB data, also Puttini et al. [24]. A Bayesian classification was used to detect anomalous in network traffic in Mobile Ad Hoc Networks (MANET) using SNMP-MIB variables.

A key observation drawn from the literature related to SNMP-MIB data-based attack detection is that most of the studies mentioned above are limited to specific types and a limited number of attacks. Also, some of these studies focused on the detection of anomalous traffic as it distinct from normal without considering the determination of the attack type. So there is still a need to exploit SNMP-MIB as a rich data source for more security in computer networks. Here, we intend to offer an effective approach with high detection and accuracy rate compared to others in detecting attacks with SNMPMIB data. Our work utilizes MIB data to classify a greater number of different attacks (TCP-SYN, UDP flood, ICMPECHO, HTTP flood, Slowpost, slowloris, Brute Force) based on machine learning techniques.

III.  PROPOSED MODEL

*A. Design*

In this section, we explain the main steps we perform in our work for network attack detection and classification using SNMP-MIB data and machine learning techniques. The overall architecture and the main idea of the working model is illustrated in Fig. 2. First, a new SNMP-MIB dataset is built from network traffic data collected from a real testbed network.

Second, the collected MIB dataset is used as input to the three classifiers, as we stated earlier, are used to build classification models on the MIB data to classify attacks by type. Towards studying the effect of using MIB data on detecting network attacks, we conduct two sets of experiments: (1) The MIB variables collected were used with an accurate feature selection method to select the most effective variables, and then machine learning classifiers were used to classify the type of attack. (2) the collected MIB variables are categorized into their corresponding groups (Interface, IP, TCP and ICMP). In this paper, the focus is only one group, namely the Interface group. The Interface group variables are listed in Table I. (3) The classifiers then are applied to the Interface group. The remainder of the sections presents our methodology in further detail.

TABLE I. INTERFACE MIB GROUP VARIABLES

| Interface MIB Group | | | |
|---|---|---|---|
| Variable 1 | ifInOctets | Variable 5 | ifInNUcastPkts |
| Variable 2 | ifOutOctets | Variable 6 | ifInDiscards |
| Variable 3 | ifoutDiscards | Variable 7 | ifOutUcastPkts |
| Variable 4 | ifInUcastPkts | Variable 8 | ifOutNUcastPkts |

### B. SNMP-MIB Dataset

In our experiments, The SNMP-MIB dataset [15] is used for testing our proposed framework. The MIB dataset contains approximately 4998 records with 34 MIB variables. The data records of attacks fall into 6 types of DoS attacks (TCP-SYN, UDP flood, ICMP-ECHO, HTTP flood, Slowloris, Slowpost) and Brute Force attack. Further description for the MIB dataset and attacks instances can be found in [15].

### C. Machine Learning Classifiers

Classification is one of the most commonly applied supervised machine learning techniques. Machine Learning Classifiers have been successfully applied to intrusion detection towards finding various and effective approaches to detect intrusion [13]. Classifiers are used to basically classify the network traffic into normal and abnormal categories. The goal is to build a model from classified objects and use the model to classify new objects as accurately as possible [1], [9]. In this work, we chose three different classification algorithms: AdaboostM1, Random Forest and Multilayer Perceptron (MLP). We know from the literature that the performance of these classifiers are highly accurate in classification problems. These classifiers were applied to our SNMP-MIB dataset in order to classify attack and normal traffic, and then evaluated for accuracy in order to investigate the ability and effectiveness of SNMP-MIB data in detecting different types of networks attacks with machine learning technique. To the best of our knowledge, the Random Forest classifier has not been utilized in SNMP-MIB-based attack detection and classification until now, while several classification algorithms, such as BP, C4.5, Bayesian networks and SVM have been used with SNMP-MIB data [1][9].

## IV. EXPERIMENTS

### A. Implementation

For the purpose of this work, we employed machine learning techniques to evaluate our MIB dataset in attacks detection and classification. For conducting the experiments with classifiers, initially, we randomly divided the MIB dataset into 70% as a training dataset (3498 records) and 30% (1500 records) as a testing dataset. Both datasets contained normal and the other seven attack classes. In this paper, we applied AdaboostM1, Random Forest and ANN (MLP) classifiers to the MIB datasets. The method used the Interface MIB variables, and then the classification algorithms are applied to the group. Our experiments with machine learning techniques are conducted using an open source data mining toolkit [25]: WEKA 3.7.3. WEKA (Waikato Environment for Knowledge Analysis) was originally developed at the University of Waikato, New

TABLE II. CONFUSION MATRIX FOR TWO CLASSES

| | | Predicted Class | |
|---|---|---|---|
| Actual Class | | Positive | Negative |
| | Positive | TP | FP |
| | Negative | FN | TN |

Zealand. This tool is written in Java language and has a collection of Machine Learning and Data Mining algorithms for data pre-processing, clustering, classification and others [25]. From WEKA, an ensemble classification method named AdaboostM1 is used with the J48 algorithm as a base classifier. Another classifier called Random Forest is also applied with the number of trees equal to 100. The third classifier used is MLP, which is a back propagation learning algorithm, with a learning rate = 0.3, momentum = 0.2 and the number of epochs equal to 500. The transfer function used is a Sigmoid function.

### B. Performance Evaluation Metrics

In our work, we used some important metrics to evaluate how accurately the machine learning classifiers are detecting and classifying the different types of attack based on the MIB dataset. Performance metrics used including Precision, Recall, F-measure and Accuracy as shown in (1) to (4). Precision is the ratio of the predicted positive samples that were correctly classified, and Recall is the ratio of positive samples that were classified correctly as positive. It refers to the true positive rate and it is also known as Sensitivity measure. F-Measure is a measure of a classification models accuracy depending on the precision and the recall metrics. It is considered as a weighted average of both Precision and Recall metrics. Confusion Matrix (Table II) also used to evaluate classifier performance and to visualize the performance of the classification model; that is demonstrated using information about actual and predicted classifications of the model used.

$$Precision = \frac{TP}{TP + FP} \quad (1)$$

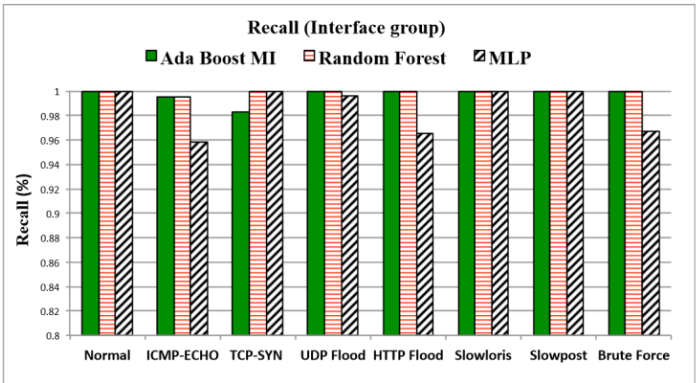

$$Recall = \frac{TP}{TP + FN} \qquad (2)$$

$$F - Measure = \frac{2 \cdot Precision \cdot Recall}{Precision + Recall} \qquad (3)$$

$$Accuracy = \frac{TP + FN}{TP + FP + TN + FN} \qquad (4)$$

Given that F-Measure is an evaluation measure, which depends on the precision and the recall metrics, hence it will be considered in all our experimental results as shown in the next section.

## V. EXPERIMENTAL RESULTS AND DISCUSSION

This section describes and explains the experimental results for the proposed method that uses the full MIB dataset (both training and testing datasets). In this paper, we used only one MIB group (the Interface MIB Group) that contains the affiliated MIB variables, and then the classification algorithms (AdaboostM1, Random Forest and MLP classifiers) are used.

Fig. 3. Recall results with Interface group.

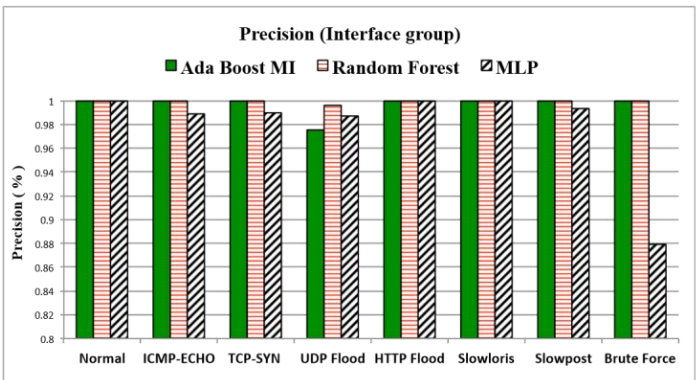

Fig. 4. Precision results with Interface group.

The aim here is to study and show the impact of the Interface group in attacks classification by means of the machine learning classifiers.

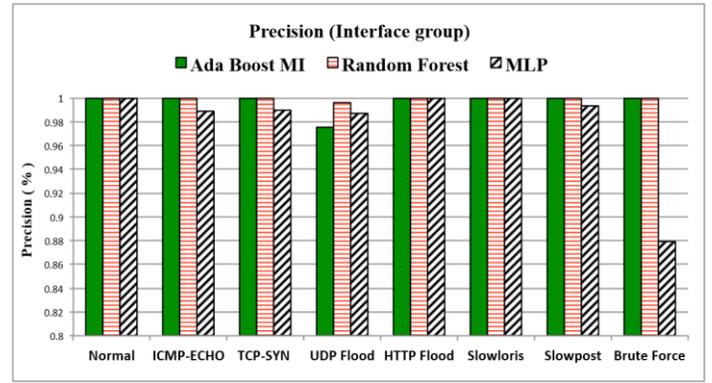

### A. Experimental Results with Interface MIB Group

In this experiment, the AdaboostM1, Random Forest and MLP classifiers are applied and evaluated on the Interface group dataset. Fig. 3, 4, and 5 show the performance of these classifiers in terms of Recall, Precision, and F-measure (respectively), based on the Interface MIB variables.

The results in Fig. 3 indicate that the three classifiers achieved very high results in correctly identifying the normal traffic records, as well as the Slowloris and Slowpost attack records in the testing set, with high recall values of 1. From the results, AdaboostM1 and Random Forest classifiers succeeded in identifying most types of attack, followed by MLP classifiers, which also achieved high performance in attack detection with the interface group. More specifically, the Random Forest classifier achieved the best recall results for all types of attack. Fig. 4 shows that precision results are very high for the three classifiers in all types of attack except for the AdaboostM1 classifier in the UDP flood attack where its performance was less effective compared to other attacks. Also, the MLP classifier has the minimum result for the Brute Force attack; this means that MLP cannot identify Brute Force attack records among all the other records. The results in Fig. 5 indicate that the three classifiers achieved very high results in correctly

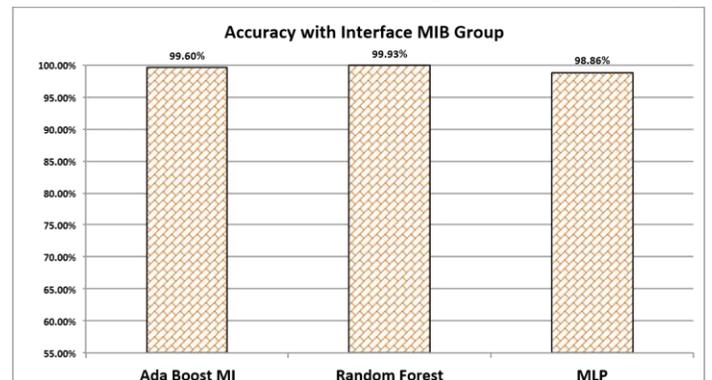

identifying the normal traffic records, as well as the

Fig. 5. F-Measure results with Interface group.

Fig. 6. Accuracy rate for the three classifiers with interface group.

Slowloris and Slowpost attack records in the testing set, with high recall values of 1. From the results, AdaboostM1 and Random Forest classifiers succeeded in identifying most types of attack, followed by MLP classifiers, which also achieved high performance in attack detection with the interface group. More specifically, the Random Forest classifier achieved the best recall results among all.

From Fig. 5, we can observe that F-Measure results are very high for the three classifiers in all types of attack except for the AdaboostM1 classifier in the UDP flood attack where its performance is less effective compared to other attacks. Also, the MLP classifier has the minimum result for the Brute Force attack, which means that MLP cannot identify Brute Force attack records among all the other records. In a summary, the F-measure results in Fig. 5 confirm high accuracy for all attacks. This means that Interface MIB variables have a significant impact in identifying precisely and correctly most types of attack by the three classifiers used.

Experimental results in terms of accuracy rate for all three classifiers with the Interface MIB group dataset are shown in Fig. 6. As depicted, the accuracy results indicate that the Random Forest classifier outperformed in classifying the intrusions (attacks) with the Interface MIB group dataset, followed by the AdaboostM1 classifier that achieved high accuracy results and very close by the Random Forest results. However, the MLP classifier shows the lowest accuracy rates with the Interface MIB group dataset compared to the other classifiers.

## VI. CONCLUSIONS

In this paper, we have presented a methodology for network attack detection based on SNMP-MIB data by applying machine learning techniques. The purpose of our work was to prove the ability and effectiveness of SNMP-MIB data in network anomaly detection by demonstrating the detection of the largest possible number of the most common and modern attacks that can occur on the Interface layer. Our methodology involved three classification algorithms, namely, AdaboostM1, Random Forest and MLP classifiers. In our approach, we used the MIB variables in the Interface MIB group, which included a number of MIB variables that were affiliated to it. The classification algorithms were then applied to the MIB group to show how the group is affected by attacks, and therefore to determine the effectiveness of the Interface MIB group in anomaly detection. From the results of this approach, we found that the performance of each classifier is different using the Interface MIB group, where the accuracy rate varied between high and low for the three classifiers. The Random Forest classifier achieved the highest accuracy rate with the Interface group 99.93%. From the results, we also concluded that all types of attack affected the Interface MIB group. The overall conclusion is that using SNMP-MIB data with machine learning techniques is a very effective approach to network anomaly detection with a distinctive effect on network security.

## REFERENCES

[1] Yu, J., Lee, H., Kim, M. S., & Park, D. (2008). Traffic flooding attack detection with SNMP MIB using SVM. Computer Communications, 31(17), 4212-4219.

[2] Moore, D., Shannon, C., Brown, D. J., Voelker, G. M., & Savage, S. (2006). Inferring internet denial-of-service activity. ACM Transactions on Computer Systems (TOCS), 24(2), 115-139.

[3] Kirnapure, W. K., & Patil, A. R. B. (2017). Classification, Detection and Prevention of Network Attacks Using Rule Based Approach.

[4] Nanda, N. B., & Parikh, A. (2017). Classification and Technical Analysis of Network Intrusion Detection Systems. International Journal of Advanced Research in Computer Science, 8(5).

[5] Gupta, J., & Singh, J. (2017). Detecting Anomaly Based Network Intrusion Using Feature Extraction and Classification Techniques. International Journal, 8(5).

[6] Bao, C. M. (2009, August). Intrusion detection based on one-class svm and snmp mib data. In Information Assurance and Security, 2009. IAS'09. Fifth International Conference on (Vol. 2, pp. 346-349). IEEE.

[7] Thottan, M., & Ji, C. (2003). Anomaly detection in IP networks. IEEE Transactions on signal processing, 51(8), 2191-2204.

[8] Wu, Q., & Shao, Z. (2005, October). Network anomaly detection using time series analysis. In Autonomic and Autonomous Systems and International Conference on Networking and Services, 2005. ICAS-ICNS 2005. Joint International Conference on (pp. 42-42). IEEE.

[9] Yu, J., Kang, H., Park, D., Bang, H. C., & Kang, D. W. (2013). An indepth analysis on traffic flooding attacks detection and system using data mining techniques. Journal of Systems Architecture, 59(10), 1005-1012.

[10] Li, J., & Manikopoulos, C. (2003, June). Early statistical anomaly intrusion detection of DOS attacks using MIB traffic parameters. In Information Assurance Workshop, 2003. IEEE Systems, Man and Cybernetics Society (pp. 53-59). IEEE.

[11] Muda, Z., Yassin, W., Sulaiman, M. N., & Udzir, N. I. (2016). KMeans Clustering and Naive Bayes Classification for Intrusion Detection. Journal of IT in Asia, 4(1), 13-25.

[12] Rahmani, C., Sharifi, M., & Tafazzoli, T. (2004, February). An Exprimental Analysis of Proactive Detection of Distributed Denial of Service Attacks. In Proceedings of the IIT Kanpur Hackers Workshop (IITKHACK04) (pp. 37-44).

[13] NAMVARASL, S., & AHMADZADEH, M. (2014). A Dynamic Flooding Attack Detection System Based on Different Classification Techniques and Using SNMP MIB Data. IJCNCS Publication, 2(9).

[14] Park, J. S., & Kim, M. S. (2008). Design and implementation of an SNMP-based traffic flooding attack detection system. Challenges for next generation network operations and service management, 380-389.

[15] Al-Kasassbeh, M., Al-Naymat, G., & Al-Hawari, E. (2016). Towards Generating Realistic SNMP-MIB Dataset for Network Anomaly Detection. International Journal of Computer Science and Information Security, 14(9), 1162.

[16] Cabrera, J. B., Lewis, L., Qin, X., Lee, W., & Mehra, R. K. (2002). Proactive intrusion detection and distributed denial of service attacksa case study in security management. Journal of Network and Systems Management, 10(2), 225-254.

[17] Hsiao, H. W., Lin, C. S., & Chang, S. Y. (2009, August). Constructing an ARP attack detection system with SNMP traffic data mining. In Proceedings of the 11th international conference on electronic commerce (pp. 341-345). ACM.

[18] Cerroni, W., Moro, G., Pirini, T., & Ramilli, M. (2013, January). Peerto-peer data mining classifiers for decentralized detection of network attacks. In Proceedings of the Twenty-Fourth Australasian Database Conference-Volume 137 (pp. 101-107). Australian Computer Society, Inc.

[19] Cerroni, W., Moro, G., Pasolini, R., & Ramilli, M. (2015). Decentralized detection of network attacks through P2P data clustering of SNMP data. Computers & Security, 52, 1-16.

[20] Priya, P. M., Akilandeswari, V., Shalinie, S. M., Lavanya, V., & Priya, M. S. (2014, April). The Protocol Independent Detection and Classification (PIDC) system for DRDoS attack. In Recent Trends in Information Technology (ICRTIT), 2014 International Conference on (pp. 1-7). IEEE.

[21] Al-Kasassbeh, M., & Adda, M. (2009). Network fault detection with Wiener filter-based agent. Journal of Network and Computer Applications, 32(4), 824-833.

[22] Al-Kasassbeh, M. (2011). Network Intrusion Detection with Wiener Filter-Based Agent. World Appl. Sci. J, 13(11), 2372-2384.

[23] Vyavhare, A., Bhosale, V., Sawant, M., & Girkar, F. (2012). Cooperative Wireless Intrusion Detection System Using MIBs From SNMP. International Journal of Network Security & Its Applications, 4(2), 147.

[24] Puttini, R., Hanashiro, M., Miziara, F., de Sousa, R., Garca-Villalba, L. J., & Barenco, C. J. (2006). On the anomaly intrusion-detection in mobile ad hoc network environments. Lecture Notes in Computer Science, 4217, 182.

[25] Hall, M., Frank, E., Holmes, G., Pfahringer, B., Reutemann, P., & Witten, I. H. (2009). The WEKA data mining software: an update. ACM SIGKDD explorations newsletter, 11(1), 10-18.

[26] Alkasassbeh, M., Al-Naymat, G., Hassanat, A., Almseidin, M. (2016). Detecting Distributed Denial of Service Attacks Using Data Mining Techniques, International Journal of Advanced Computer Science and application, Vol. 7, Issue 1, pp. 436-445, January 2016.